\documentclass[slac_one]{revtex4}
\usepackage{graphicx}
\usepackage{fancyhdr}
\def\lN{l_{\rm n}}
\def\lF{l_{\rm f}}

\def\low{{\rm low}}
\def\high{{\rm high}}
\def\equal{{\rm eq}}

\def\qL{{\tilde q_L}}

\def\mll{m_{ll}}

\def\mqlN{m_{q\lN}}
\def\mqlF{m_{q\lF}}

\def\mqlLow{m_{ql(\low)}}
\def\mqlHigh{m_{ql(\high)}}

\def\NO{{\tilde\chi^0_1}} 
\def\NT{{\tilde\chi^0_2}} 
\def\lR{{\tilde l_R} }

\def\mll{m_{ll}}
\def\max{{\rm max}}

\def\mNO{m_{\NO}}
\def\mlR{m_{\lR}}
\def\mNT{m_{\NT}}
\def\mqL{m_{\qL}}

\def\sNO{{m_\NO^2}}
\def\slR{{m_\lR^2}}
\def\sNT{{m_\NT^2}}
\def\sqL{{m_\qL^2}}

\def\maxmll{m^\max_{ll}}
\def\maxmqll{m^\max_{qll}}
\def\maxmqlLow{m^\max_{ql(\low)}}
\def\maxmqlHigh{m^\max_{ql(\high)}}

\def\itB{\it}
\def\LSfun{\chi^2}
\def\SPSOaa{SPS~1a~${(\alpha)}$}
\def\SPSOab{SPS~1a~${(\beta)}$}

\def\pbeta{($\beta$)}
\def\av#1{\langle #1 \rangle}
\def\spcA{\mbox{\hspace{1.2ex}}}

\begin{document}

\title{{\small{2005 International Linear Collider Workshop - Stanford,
U.S.A.}}\\ 
\hfill {\rm\small CERN-PH-TH/2005-120\\[-.5ex] \ \hfill
\\}%
\vspace{6pt}
Resolving ambiguities in mass determinations
at future colliders} 

%

\author{B.~K.~Gjelsten}
\affiliation{Department of Physics, 
University of Oslo, 
N-0316 Oslo, Norway 
and Laboratory for High Energy Physics, University of Bern, CH~3012 Bern, Switzerland}
\author{D.~J.~Miller}
\affiliation{Department of Physics and Astronomy, 
University of Glasgow, 
Glasgow G12 8QQ, 
U.K.}
\author{P.~Osland}
\affiliation{Department of Physics and Technology, 
University of Bergen, N-5007 Bergen, Norway
and TH Division, Physics Department, CERN, CH~1211 Geneva, Switzerland}

\begin{abstract}
The measurements of kinematical endpoints, in cascade decays 
of supersymmetric particles, in principle allow for a determination
of the masses of the unstable particles.
However, in this procedure ambiguities often arise.
We here illustrate how such ambiguities arise.
They can be resolved by a precise determination of the LSP mass,
provided by the Linear Collider.
\end{abstract}

\maketitle

\thispagestyle{fancy}


\section{INTRODUCTION} 
If R-parity conserving supersymmetry exists below the 
TeV-scale, new particles will be produced and decay in cascades 
at the LHC. The lightest supersymmetric particle will escape the 
detectors, thereby complicating the full reconstruction of the 
decay chains. 
However, the masses of the particles in a cascade like
\begin{equation}
\tilde q_L \to \NT q \to \lR^\mp \lN^{\pm} q \to \NO \lF^{\mp} \lN^{\pm} q
\label{eq:cascade}
\end{equation}
can be determined
from endpoints of kinematical distributions 
\cite{Baer:1995va,Allanach:2000kt,Gjelsten:2004ki}. 
Here, $\tilde q_L$ denotes a (left-handed) squark,
$\lR$ a (right-handed) slepton, whereas $\NT$ and $\NO$ denote 
the lightest neutralinos, the latter being stable.
The observed leptons and quark (jet) are denoted $\lN$, $\lF$ and $q$,
where the subscripts ``n'' and ``f'' denote ``near'' and ``far''.

Attention is focused on the mSUGRA benchmark point SPS~1a (line)
\cite{Allanach:2002nj}
\begin{equation}  
m_0=-A_0=0.4\, m_{1/2}, \quad
\tan\beta=10, \quad \mu>0,
\label{eq:slopedefinition}
\end{equation}  
and in particular two points on the line, SPS~1a~($\alpha$) 
with $m_0=100~\text{GeV}$ and $m_{1/2}=250~\text{GeV}$,
and SPS~1a~($\beta$) with 
$m_0=160~\text{GeV}$ and $m_{1/2}=400~\text{GeV}$.
The resulting low-energy masses entering in the cascade (\ref{eq:cascade})
are given in Table~\ref{sps1a-masses}
\cite{Gjelsten:2004ki}.

\begin{table}[htb]
\begin{center}
\caption{SPS 1a masses}
\begin{tabular}{|l|c|c|c|c|}
\hline 
& $\mqL$ [GeV] & $\mNT$ [GeV] & $\mlR$ [GeV] & $\mNO$ [GeV] \\
\hline 
($\alpha$) & 537.2 & 176.8 & 143.0 & 96.1 \\
($\beta$) & 826.3 & 299.1 & 221.9 & 161.0 \\
\hline
\end{tabular}
\label{sps1a-masses}
\end{center}
\end{table}

Invariant masses involving various subsets of the observed particles, namely
the quark, $q$ and the two leptons, $\lN$ and $\lF$, can be reconstructed.
Since the ``near'' and ``far'' leptons can not be distinguished, one must
instead, on an event-by-event basis, define a ``high'' and ``low''
distribution, such that
\begin{equation}
\mqlLow=\min(\mqlN,\mqlF), \quad
\mqlHigh=\max(\mqlN,\mqlF).
\end{equation}
From a knowledge of kinematical endpoints, in particular those of the
$m_{qll}$, $\mqlLow$, $\mqlHigh$ and $\mll$ distributions,
the masses of the unstable particles can be reconstructed.
Indeed, the endpoints of these distributions can be expressed
explicitly in terms of $\mqL$, $\mNT$, $\mlR$ and $\mNO$
\cite{Allanach:2000kt,Gjelsten:2004ki}.
Additional information can be obtained from threshold determinations 
\cite{Allanach:2000kt}, and the method can be extended to
the case of a gluino at the head of the chain \cite{Gjelsten:2005aw}.

\section{COMPOSITE FORMULAS}
Many of the problems which arise in this method are related to the
fact that the kinematical endpoints are composite functions of the
unknown masses.  Indeed, the functional form for $\maxmqll$, $\maxmqlLow$
and $\maxmqlHigh$ depend on the relative mass ratios.  For $\maxmqlLow$ and
$\maxmqlHigh$, they are given by \cite{Allanach:2000kt,Gjelsten:2004ki}:
\begin{equation}
\big(\maxmqlLow,\maxmqlHigh\big) 
= \left\{
\begin{array}{llcc}
\big(m^\max_{q\lN},m^\max_{q\lF}\big) &
\quad {\rm for } \quad
& 2\slR > \sNO+\sNT > 2\mNO\mNT & \quad {\itB (1)} \\[4mm]
\big(m^\max_{ql(\equal)},m^\max_{q\lF}\big) &
\quad {\rm for } \quad
& \sNO+\sNT > 2\slR > 2\mNO\mNT & \quad {\itB (2)}\\[4mm]
\big(m^\max_{ql(\equal)},m^\max_{q\lN}\big) &
\quad {\rm for } \quad
& \sNO+\sNT > 2\mNO\mNT > 2\slR & \quad {\itB (3)}
\end{array} \right\} 
\label{eq:m_ql}
\end{equation}
with
\def\sqL{{m_\qL^2}}
\begin{eqnarray}
\big(m^\max_{q \lN}\big)^2
&=& \big(\sqL-\sNT\big)\big(\sNT-\slR\big)/\sNT
\label{Eq:edge-qlN} \\[4mm]
\left(m^\max_{q \lF}\right)^2
&=& \big(\sqL-\sNT\big)\big(\slR-\sNO\big)/\slR
\label{Eq:edge-qlF} \\[4mm]
\big(m^\max_{ql(\equal)}\big)^2 &=&
\big(\sqL-\sNT\big)\big(\slR-\sNO\big)/\big(2\slR-\sNO\big)
\label{Eq:edge-ql-equal} 
\end{eqnarray}

In this report we will focus on $\maxmqlLow$ and $\maxmqlHigh$ as functions
of $\mNO$, since the behaviour of these two endpoints is most important 
for the mSUGRA points studied. 
In the case of $\maxmqlLow$, the functional form changes when
$\mNO$ crosses $\sqrt{2\mlR^2-\mNT^2}$, whereas for $\maxmqlHigh$ it
changes when $\mNO$ crosses $\mlR^2/\mNT$.  The three cases given in
eq.~(\ref{eq:m_ql}) are in \cite{Gjelsten:2004ki} referred to as
regions {\itB(1,1)}, {\itB(1,2)} and {\itB(1,3)}, where the first
index (``{\itB1}'') refers to the expression for $\maxmqll$, which 
remains unchanged.  The corresponding critical mass values are given in
Table~\ref{tab:regions} for $\mNO$, keeping the other masses at their
nominal SPS~1a values.

\begin{table}[htb]
\begin{center}
\caption{Region borders for $\mNO$ [GeV]}
\begin{tabular}{|l|c|c|c|}
\hline 
& SPS 1a nominal & Region {\itB(1,1)} vs.\ {\itB(1,2)} 
& Region {\itB(1,2)} vs.\ {\itB(1,3)}  \\
\hline 
($\alpha$) & 96.1 & 98.2 & 115.7 \\
($\beta$) & 161.0  & 95.0 & 164.6 \\
\hline
\end{tabular}
\label{tab:regions}
\end{center}
\end{table}

\begin{figure*}[htb]
\centering
\includegraphics[width=75mm]{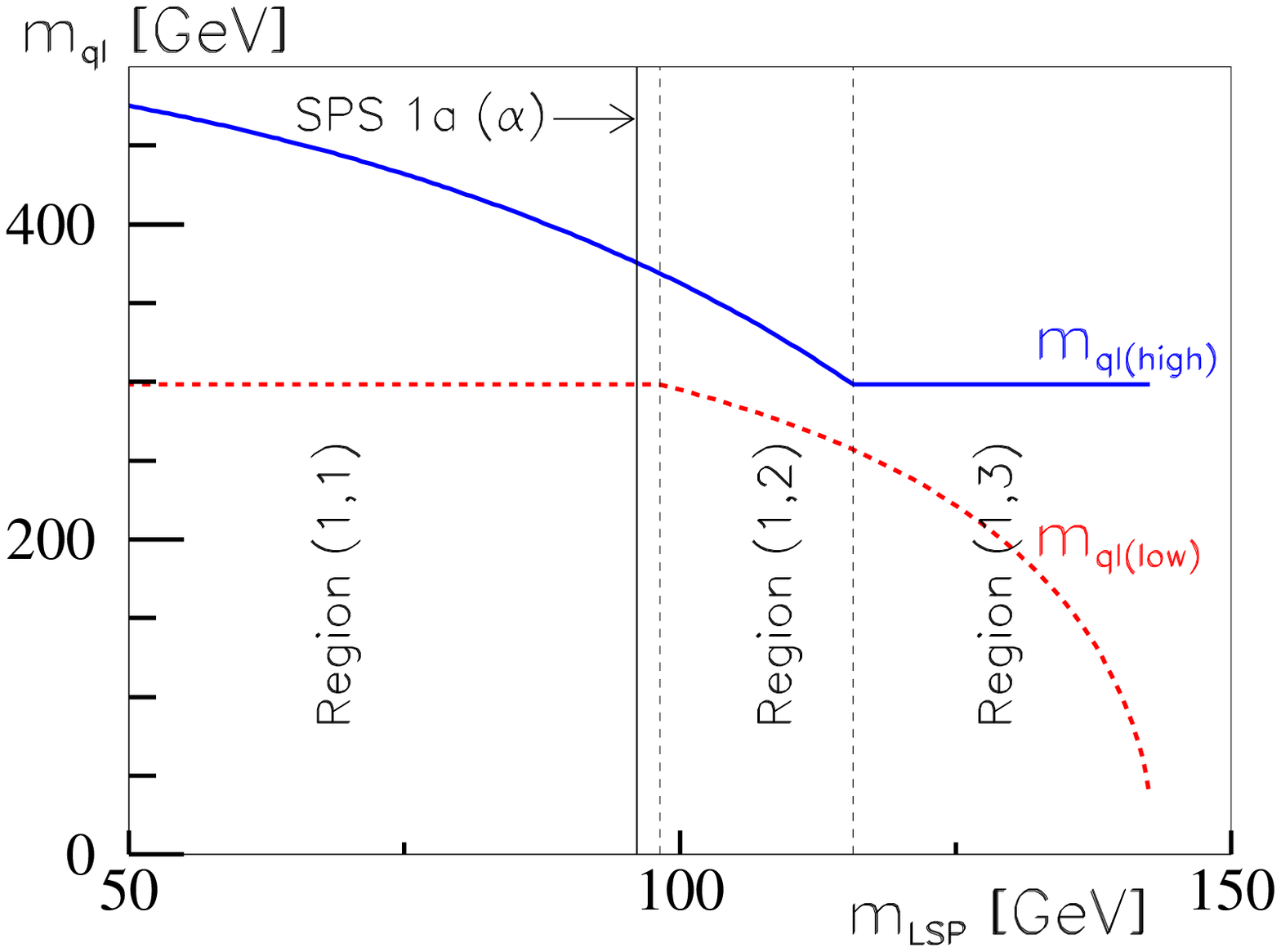}
\includegraphics[width=75mm]{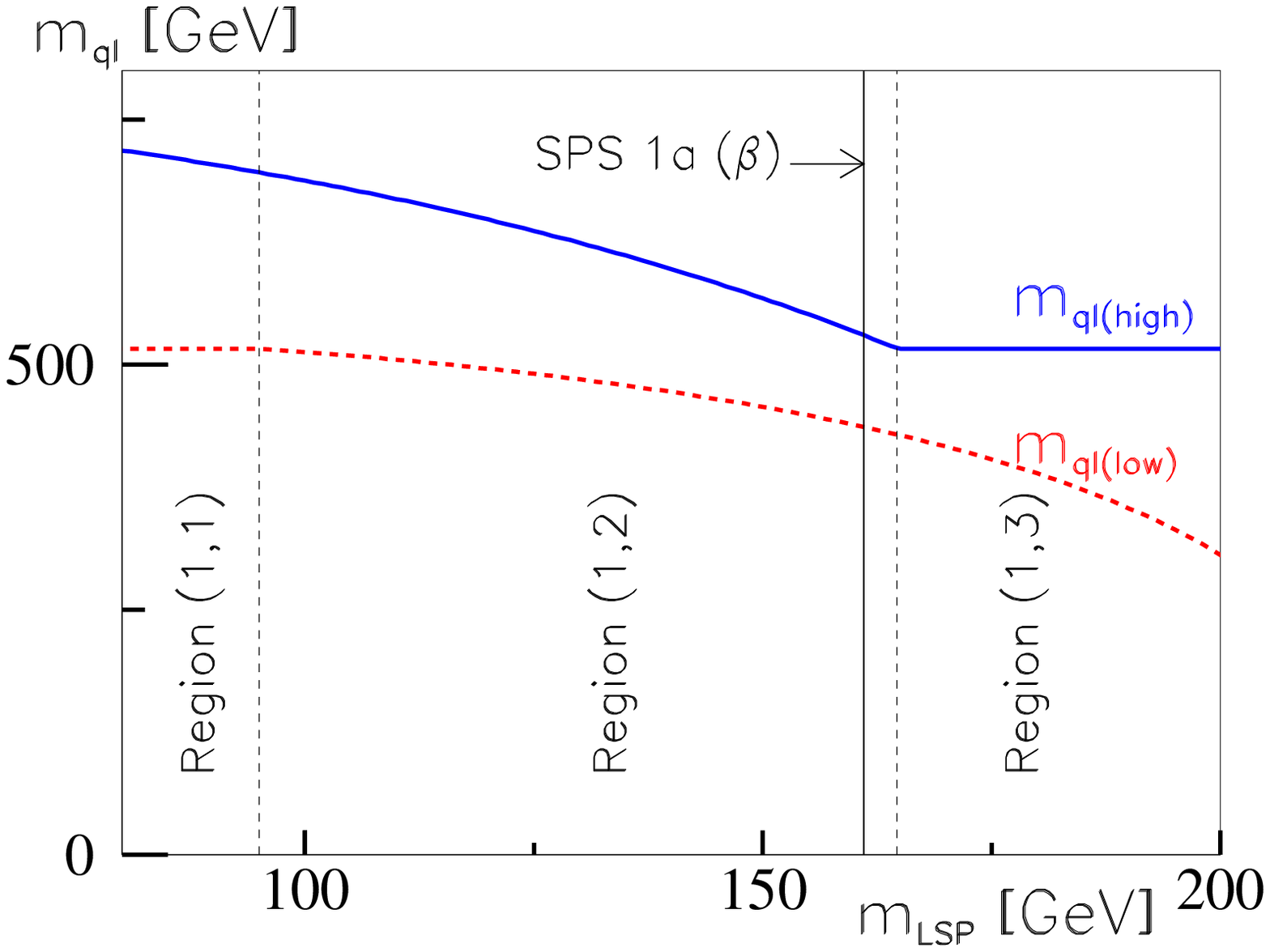}
\caption{The invariants $\maxmqlLow$ and $\maxmqlHigh$ are
composite functions of $\mNO$ (denoted $m_{\rm LSP}$). 
Other masses are held fixed at respectively the SPS~1a ($\alpha$) (left panel) 
and SPS~1a ($\beta$) (right panel) values.} \label{fig:m-ql}
\end{figure*}

Thus, the functional forms for $\maxmqlLow$ and $\maxmqlHigh$ change very close
to the nominal values of $\mNO$ for SPS~1a~($\alpha$)
and ($\beta$), respectively, as is illustrated in Fig.~\ref{fig:m-ql}.
These points are close to the transitions from region {\itB(1,1)} to
{\itB(1,2)}, and from region {\itB(1,2)} to {\itB(1,3)}, respectively.

\section{AMBIGUITIES}

Ambiguities in the masses extracted from endpoint measurements are
principally caused by the analytic form of the endpoint expressions,
as seen above. In particular, they frequently lead to multiple solutions in
mass-space for a particular set of endpoint values, even when experimental
uncertainties are neglected. Furthermore, inclusion of 
extra overconstraining measurements moves solutions about in mass-space 
and can destroy or create new ones, adding to the complexity. 

\subsection{Multiple solutions}
\label{sec:multiple}

\begin{figure*}[htb]
\centering
\includegraphics[width=90mm]{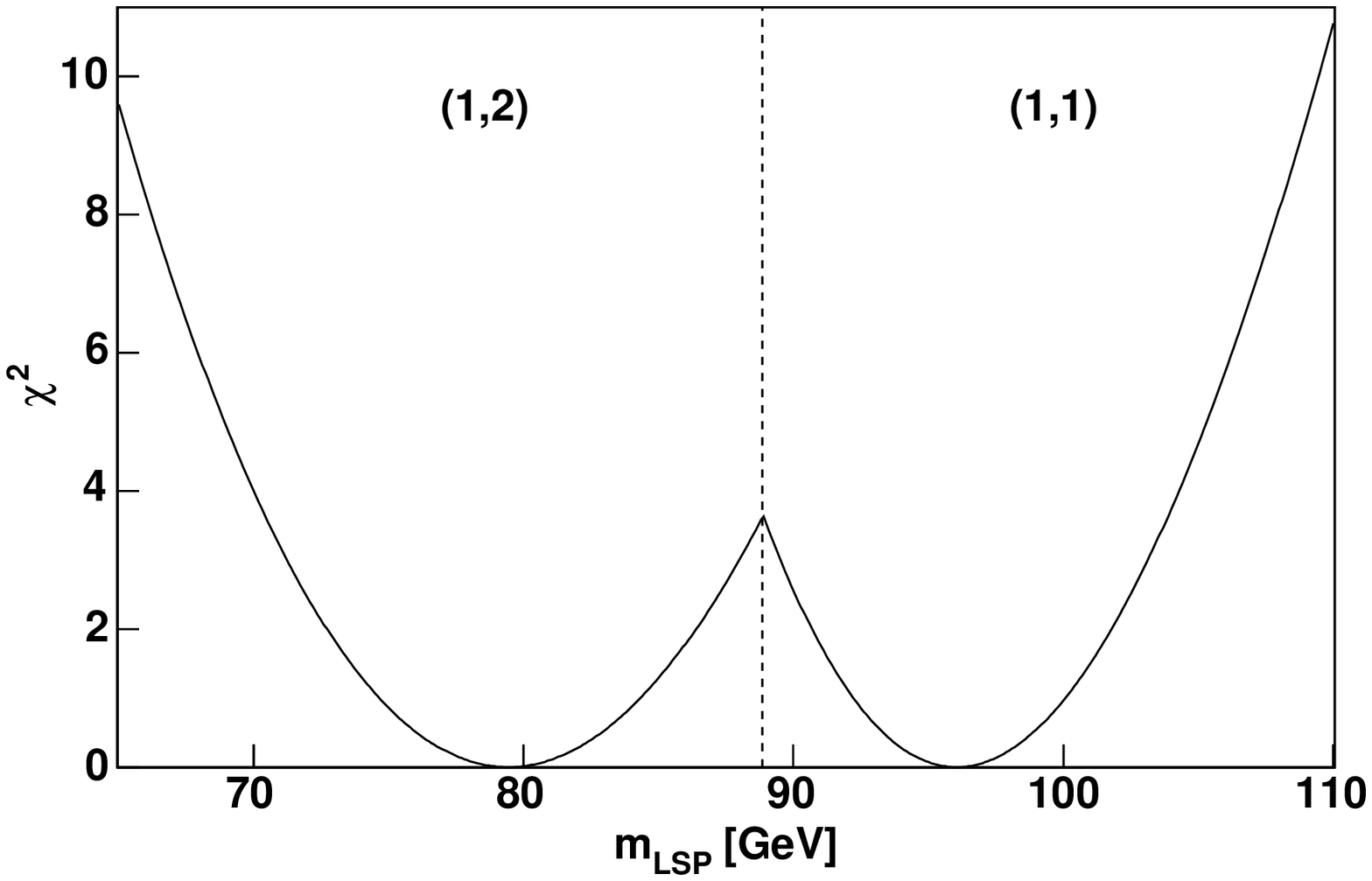} 
\hspace{-5mm}
\includegraphics[width=90mm]{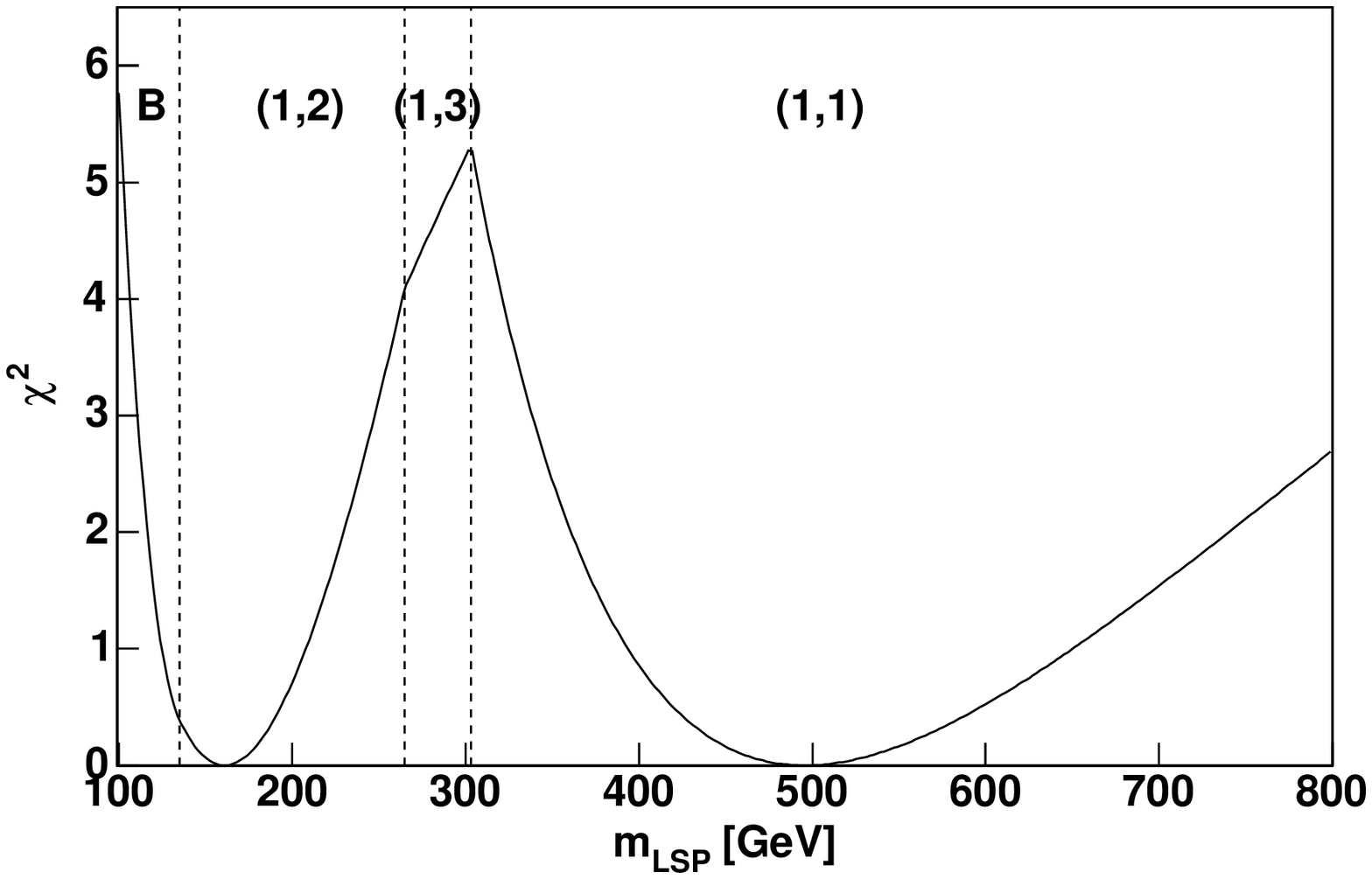}
\caption{$\chi^2$ vs.\ $\mNO$,
allowing the other masses to vary.
Left: SPS~1a ($\alpha$), 
Right: SPS 1a ($\beta$).} \label{fig:scans}
\end{figure*}

In this subsection we restrict the discussion to the case in which the 
number of available (linearly independent) endpoints is equal to the 
number of masses involved. In particular we have in mind the standard 
case where $\maxmll$, $\maxmqlLow$, $\maxmqlHigh$ and $\maxmqll$ 
are measured. 
In this situation analytic expressions for the masses in terms of the 
endpoints can be obtained \cite{Gjelsten:2004ki}, which in a numerical 
fit would correspond to solutions with $\chi^2=0$. 
While the endpoints are obviously single-valued functions of the masses, 
 due to the composite (and quadratic) form of the endpoint expressions 
the inverse is in general not true. 
A specific set of endpoint values can often be produced by several sets 
of masses. This ambiguity is illustrated in Fig.~\ref{fig:scans}, where 
multiple solutions in the
$\chi^2$ fit become evident.  In this fit, the four kinematical
endpoints $\maxmll$, $\maxmqlLow$, $\maxmqlHigh$ and $\maxmqll$ are taken at
their nominal values, and for each value of $\mNO$ in the figure, the
other three masses are allowed to vary.  The left panel is devoted to
SPS~1a~($\alpha$), whereas the right panel is devoted to
SPS~1a~($\beta$). 
Three mass regions 
are realised; {\it(1,1)}, {\it(1,2)} and {\it(1,3)}, as well as the border 
between {\it(1,2)} and {\it(1,3)}, denoted ``B''. 
Vertical lines separate the regions in the $m_{\rm LSP}$ scans.

We clearly see that for four endpoint measurements, multiple mass sets 
provide the same endpoints 
and create an ambiguity at both the investigated mSUGRA points.
For each of the two plots the non-smooth parts of the curve correspond to 
discontinuities in the mass-space trajectory defined by the (ordered) 
collection of best-fit masses found from the scan. 
At these points two identically good minima exist and a jump is made 
from one to the other. 
While the scanned mass $\mNO$ obviously changes smoothly, the other 
masses have discontinuities at these points. For the two investigated 
mSUGRA scenarios and in particular \SPSOab, $\mlR$ and $\mqL$, more so 
than $\mNT$, change considerably at these discontinuities, the first 
by roughly 5\%. 

\subsection{Overconstrained systems}
\label{sec:overconstrained}

When an extra endpoint measurement, such as the $qll$ ``threshold'',
is introduced, the system becomes overconstrained, with five
measurements determining only four masses. One might expect this to
alleviate the ambiguity from multiple solutions seen in the previous
section, since the extra endpoint should `pick out' one of the
solutions. However, this is not always the case: the uncertainty of
endpoint measurements and the introduction of an overconstraining
measurement will move the solutions around in mass-space, possibly 
creating new solutions, or destroying old ones. This may then cause 
a further ambiguity, additional to that seen in \ref{sec:multiple}, 
and actually does so for \SPSOab. We will demonstrate the
principles involved in the creation of new solutions by using a
simplified case where only one mass is unknown and two endpoints are
measured.

\begin{figure*}[htb]
\centering
\includegraphics[width=75mm]{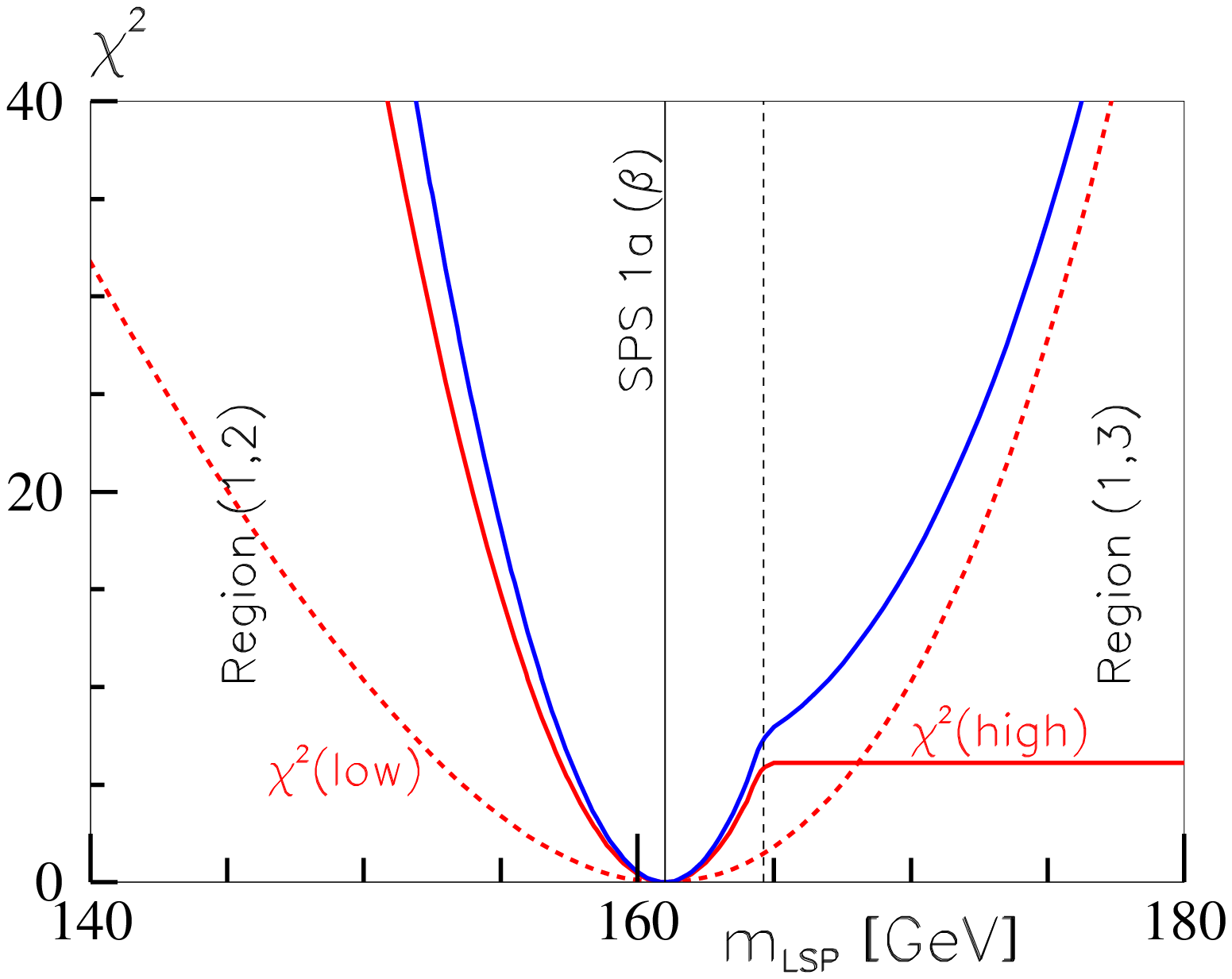}
\includegraphics[width=75mm]{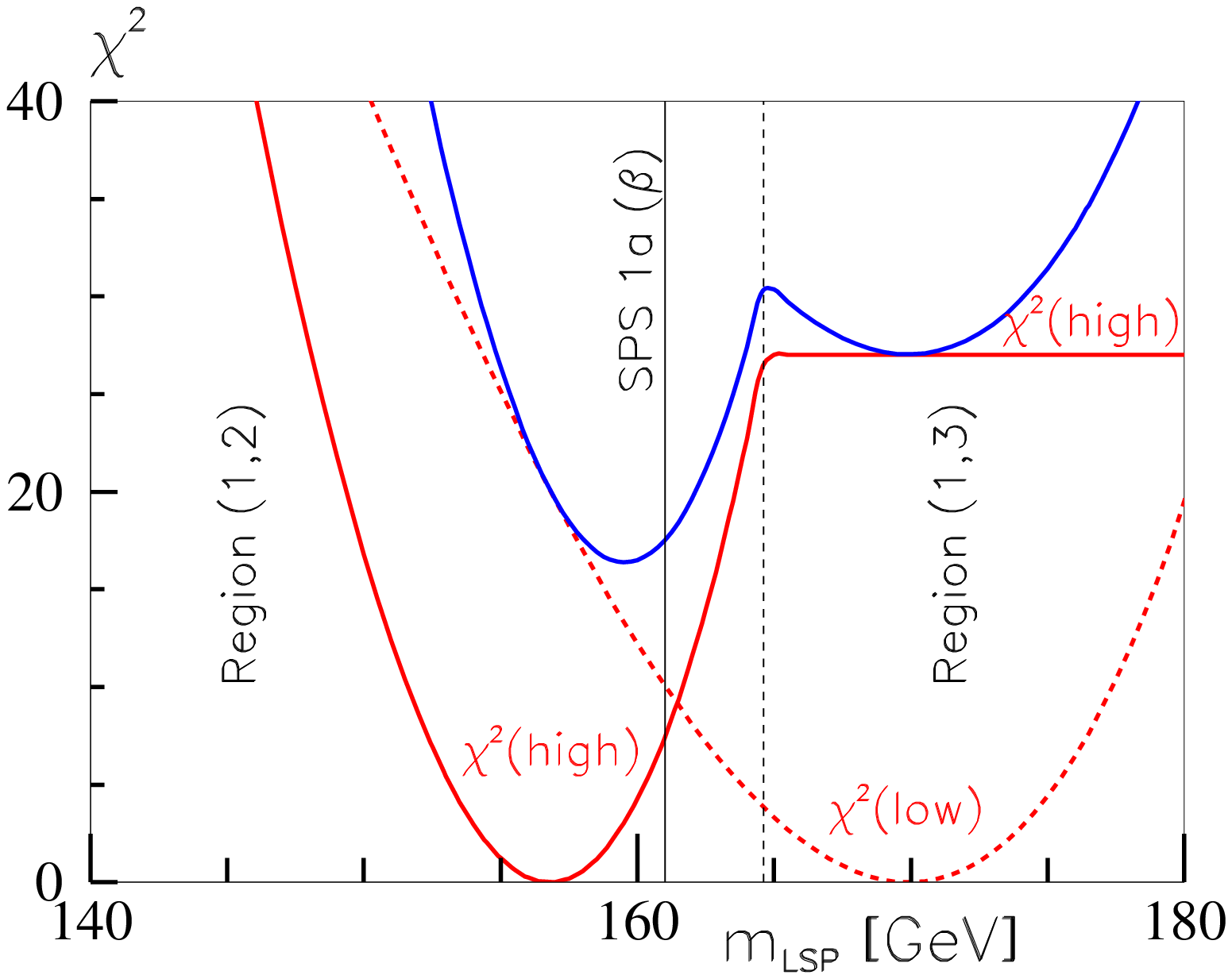}
\caption{$\chi^2$ vs.\ $\mNO$ (denoted $m_{\rm LSP}$).  The other masses are
held fixed.  Left: nominal endpoint values; right: endpoint values off-set by
$-20$ and $+15$~GeV.  } \label{fig:chi2}
\end{figure*}

For our simplified example, let us keep all masses other than $\mNO$
fixed at their nominal values, but let the two endpoint values be offset
from their nominal values, as is easily imagined due to statistical
fluctuations.
We show in Fig.~\ref{fig:chi2} a simplified $\chi^2$ function:
\begin{equation}
\chi^2=a(\maxmqlLow-\maxmqlLow{}_{\rm exp})^2
+b(\maxmqlHigh-\maxmqlHigh{}_{\rm exp})^2
\end{equation}
for SPS~1a~($\beta$), where two cases of ``experimental'' data are
considered, nominal values (left panel) and off-set values (right
panel).  The coefficients $a$ and $b$ represent the experimental
uncertainties, $\sigma(\maxmqlLow)=6.3$~GeV, and
$\sigma(\maxmqlHigh)=5.5$~GeV \cite{Gjelsten:2004ki}.  Individual
$\chi^2$-values from $\maxmqlLow$ and $\maxmqlHigh$ are shown (labeled
$\chi^2(\low)$ and $\chi^2(\high)$), together with the sum.  
In this example, there is only one solution if the endpoints take their
nominal values, despite the system being overconstrained and in spite
of the compositeness of $\chi^2$ (see left panel).

However, if the endpoint measurements give values which differ from the
nominal ones, as will in general be the case, 
the situation can be more complicated, as is shown in
the right panel of Fig.~\ref{fig:chi2}.  Here, for the purpose of
illustration, we consider $\maxmqlLow$ and $\maxmqlHigh$ offset from their
nominal values by $-20$~GeV and $+15$~GeV, respectively.  A secondary
minimum has now emerged, at a higher value of $\mNO$.

If one allows also the other masses ($\mqL$, $\mNT$, $\mlR$) 
to vary,  the transition between regions {\itB(1,2)} and {\itB(1,3)} will move,
and secondly, the minima will get deeper.
In a more realistic analysis, however, the additional measurements
of $\maxmqll$ and, in particular $\maxmll$, will disfavor the second minimum.
Indeed, with the additional information of
the $qll$ ``threshold'', 
this ambiguity was observed in the more
detailed analysis reported in \cite{Gjelsten:2004ki}.  In
Tables~\ref{table:minima-SPS1a-alpha}--\ref{table:minima-SPS1a-beta} 
we report on the probability of
having false minima, and corresponding mass values, for
\SPSOaa\ and \pbeta.
\begin{table}[htb]{ 
\caption{\SPSOaa \ ambiguities \cite{Gjelsten:2004ki}.
Left: Number of minima for $\Delta\LSfun<1$ (3) and their regions.
Right: Minima for $\Delta\LSfun\leq1$ in regions {\itB(1,1)}
and {\itB(1,2)}.
Ensemble means, $\av{m}$, and root-mean-square distances
from the mean, $\sigma$, in GeV.\label{table:minima-SPS1a-alpha}} }
\begin{tabular}{|l|c|cc|}
\hline
  & $\#$ Minima & {\itB(1,1)} & {\itB(1,2)}  \\
\hline
$\Delta\LSfun\leq 1$ & 1.12 & 94\%  & 17\% \\
$\Delta\LSfun\leq 3$ & 1.30 & 97\%  & 33\% \\
\hline
\end{tabular} \hspace{5mm}
\begin{tabular}{|c|r|rr|rr|}
\hline
&  &\multicolumn{2}{|c|}{\itB(1,1)} &\multicolumn{2}{|c|} {\itB(1,2)}  \\
& Nom & $\av{m}$ & $\sigma\hspace{1.5ex}$ & $\av{m}$ 
& $\sigma \hspace{1.5ex}$  \\
\hline
$\mNO    $ &   96.1\spcA  &   96.3\spcA &  3.8\spcA &   85.3 &  3.4\spcA \\
$\mlR    $ &  143.0\spcA  &  143.2\spcA &  3.8\spcA &  130.4 &  3.7\spcA \\
$\mNT    $ &  176.8\spcA  &  177.0\spcA &  3.7\spcA &  165.5 &  3.4\spcA \\
$\mqL    $ &  537.2\spcA  &  537.5\spcA &  6.0\spcA &  523.5 &  5.0\spcA  \\
\hline
\end{tabular}
\end{table}

\begin{table}[htb]{ 
\caption{Same as Table~\ref{table:minima-SPS1a-alpha} but now for \SPSOab.
Without the $qll$ ``threshold'' two solutions were available, the nominal
{\it(1,2)} low-mass one and a {\it(1,1)} high-mass one, see
Fig.~\ref{fig:scans} (right panel).  
Due to the proximity of \SPSOab\ to the border between {\it(1,2)} and
{\it(1,3)}, with the inclusion of the threshold measurement a third solution
has emerged.  In the tables the high-mass solution (``hms'') is shown
separately as it will probably be possible to discard it from other
measurements. In the low-mass region there is either one solution, which can
also lie on the border (``B''), or two.\label{table:minima-SPS1a-beta} } }
\begin{tabular}{|l|c|c|c|c|}
\hline
  &           &  hms  &       1 solution   & 2 solutions        \\
\cline{3-5}
  & $\#$ Min & {\itB(1,1)}  & {\itB(1,2)}/{\itB(1,3)}/B & {\itB(1,2)}\&{\itB(1,3)}  \\
\hline
$\Delta\LSfun\leq 1$  &  1.19   &    5\%   &   82\%  &   16\% \\ 
$\Delta\LSfun\leq 3$  &  1.41   &   13\%   &   72\%  &   28\% \\ 
\hline
\end{tabular}
\hspace{5mm}
\begin{tabular}{|c|c|rr|rl|cc|cc|}
\hline
\multicolumn{2}{|c|}{}  & \multicolumn{2}{|c|}{hms}  & \multicolumn{2}{|c|}{1 solution} 
&   \multicolumn{4}{|c|}{2 solutions}     \\
\cline{3-10}
\multicolumn{2}{|c|}{}  & \multicolumn{2}{|c|}{{\itB(1,1)}}  
& \multicolumn{2}{|c|}{{\itB(1,2)/(1,3)}/B}  
& \multicolumn{2}{|c|}{{\itB(1,2)}} & \multicolumn{2}{|c|}{{\itB(1,3)}} \\
\hline
& Nom &$\av{m}$&$\sigma$  & ~~~~$\av{m}$ &$~~\sigma$  &$\av{m}$&$\sigma$&$\av{m}$&$\sigma$ \\
\hline
$\mNO    $ &161 & 438&  88& 175~&  ~35& 161&  22&  166&  27  \\
$\mlR    $ &222 & 518&  85& 236~&  ~37& 221&  24&  223&  28  \\
$\mNT    $ &299 & 579&  85& 313~&  ~35& 299&  22&  304&  27  \\
$\mqL    $ &826 &1146& 104& 843~&  ~44& 826&  30&  835&  36  \\
\hline
\end{tabular}
\end{table}

\section{LINEAR COLLIDER INPUT}

With Linear Collider input on the LSP mass, taken to be determined with a
precision of 0.05~GeV \cite{Aguilar-Saavedra:2001rg}, the ambiguities are
mostly resolved.  For \SPSOaa\ the false solution is removed altogether.  The
masses and their errors are given in Table~\ref{table:LHC+LC alpha}.  For
\SPSOab\ the high-mass solution no longer appears, but there is still an
ambiguity in the low-mass region, see Table~\ref{table:LHC+LC beta}.  These
masses are however very close.  Furthermore, as a Linear Collider will
measure the slepton mass with an accuracy similar to that of the LSP mass
\cite{Aguilar-Saavedra:2001rg}, one may expect the remaining ambiguity to
vanish as well.  This possibility was not considered in the analyses
\cite{Gjelsten:2004ki,Gjelsten:2005aw} and hence is not reflected in
Tables~\ref{table:LHC+LC alpha}--\ref{table:LHC+LC beta}.

\begin{table}
\caption{\SPSOaa: Effect of adding a LC measurement of the LSP
mass \cite{Gjelsten:2004ki}.\label{table:LHC+LC alpha}}
\begin{tabular}{|c|r|rr|}
\hline
\multicolumn{1}{|c|}{} & \multicolumn{1}{|c|}{} & \multicolumn{2}{|c|}{\itB(1,1)} \\
  & Nom & $\av{m}\ $ & $\sigma\hspace{1ex} $ \\
\hline
$\mNO    $ &   96.05  &   96.05 &  0.05  \\
$\mlR    $ &  142.97  &  142.97 &  0.29  \\
$\mNT    $ &  176.82  &  176.82 &  0.17  \\
$\mqL    $ &  537.25  &  537.2\spcA  &  2.5\spcA  \\  
\hline
\end{tabular}
\end{table}

\begin{table}
\caption{\SPSOab: Effect of adding a LC measurement of the LSP
mass.\label{table:LHC+LC beta}}
\begin{tabular}{|l|c|c|c|}
\hline
  &           &       1 solution   & 2 solutions  \\
  & $\#$ Min & {\itB(1,2)}/{\itB(1,3)}/B &{\itB(1,2)}\&{\itB(1,3)}  \\
\hline
 $\Delta\LSfun\leq 1$  &  1.21 &   79\% &   21\% \\
 $\Delta\LSfun\leq 3$  &  1.45 &   55\% &   45\% \\
\hline
\end{tabular}
\hspace{5mm}
\begin{tabular}{|c|c|cc|cc|cc|}
\hline
\multicolumn{2}{|c|}{} & \multicolumn{2}{|c|}{1 solution} 
& \multicolumn{4}{|c|}{2 solutions} \\
\cline{3-8}
\multicolumn{2}{|c|}{} 
& \multicolumn{2}{|c|}{{\itB(1,2)/(1,3)}/B}
& \multicolumn{2}{|c|}{\itB(1,2)} & \multicolumn{2}{|c|}{\itB(1,3)} \\
\hline
  & Nom &$\av{m}$&$\sigma$&$\av{m}$&$\sigma$&$\av{m}$&$\sigma$ \\
\hline
$\mNO    $ &161.02 & 161.02 &   0.05 & 161.02 &   0.05 & 161.02 &   0.05  \\
$\mlR    $ &221.86 & 221.15 &   3.26 & 222.22 &   1.32 & 217.48 &   1.01  \\
$\mNT    $ &299.05 & 299.15 &   0.57 & 299.11 &   0.53 & 299.05 &   0.52  \\
$\mqL    $ &826.29 & 826.1\spcA  &   6.3\spcA  & 825.9\spcA  &   5.8\spcA  & 828.6\spcA  &   5.5\spcA   \\ 
\hline
\end{tabular}
\end{table}

A second effect, not discussed above, is very clear by comparison of the 
errors on the masses obtained with and without the LC measurement: 
Fixing the LSP mass strongly reduces the errors on all the masses. 
This comes from the fact that the endpoint method determines mass 
{\it differences} much better than the actual masses themselves, 
a feature due to the way masses enter the endpoint expressions; in 
terms of differences of masses (squared). 
Actually, the errors on the masses obtained when the LHC and the LC 
are combined, Tables~\ref{table:LHC+LC alpha}--\ref{table:LHC+LC beta}, 
are very close to the errors on the mass differences 
as obtained by the LHC alone. 

The removal of ambiguities, and the higher precision are crucial for an
extrapolation to the GUT scale \cite{Allanach:2004ud},
a major goal of studying the spectroscopy of supersymmetric particles
at future accelerators.

\begin{acknowledgments}
This research has been supported in part by the Research Council of Norway.
\end{acknowledgments}

\end{document}